# Spontaneous decay rate of an excited molecule placed near a circular aperture in a metal film

Vasily V. Klimov [1, 2, 3,] *, Ilya V. Treshin [1, 2], Dmitry V. Guzatov [4]

[1]P.N. Lebedev Physical Institute, Russian Academy of Sciences, 53 Leninsky Prospekt, Moscow 119991, Russia

[2]Dukhov Research Institute of Automatics (VNIIA), 22 Sushchevskaya Street, Moscow 127055, Russia

[3]National Research Nuclear University MEPhI (Moscow Engineering Physics Institute), 31 Kashirskoye Shosse, Moscow 115409, Russia

[4]Yanka Kupala State University of Grodno, 22 Ozheshko Street, Grodno 230023, Belarus

*klimov@vniia.ru

Abstract

We have investigated the spontaneous decay rate of an excited molecule placed near a circular aperture in a metal film of finite thickness and finite conductivity. We have considered the metal film both suspended freely in vacuum and lying on a substrate. A significant effect of molecule the position and the presence of the substrate on the rate of spontaneous emission of the molecule is shown. The asymptotes which can be used to describe this process are found. Total, radiative, and non-radiative spontaneous decay rates of the excited molecule are extracted and compared. The results may be useful in the development and interpretation of experiments investigating a single molecule with a scanning optical microscope and in the design of optical nanodevices based on the control of elementary quantum systems emission with a nanohole.

## I. INTRODUCTION

Holes of different shapes in a metal film are widely used in optical and plasmonic nanodevices [1 – 3]. It is a basic element of a near-field scanning optical microscope [4 – 8], molecular sensors with nanoaperture [9 – 17], the experiment on single photon transport by a moving atom [18, 19], and a new type of a plasmon-atomic two-dimensional metamaterial [20]. Now, methods for precise positioning of a molecule inside a nanohole are being developed [21] opening new possibilities to control the radiation of a single quantum emitter.

Despite a considerable number of good experiments of fluorescence of single molecules near a single aperture or an array of holes, as far as authors know, there are no elaborated theoretical models for such problems, and interpretation of results is not a simple task. In experiments, a hole plays a dual role. It modifies both the excitation field and the rate of spontaneous emission of the molecule.

The problem of modification of spontaneous emission rate by a nanohole (Purcell effect [22]) is more complicated in comparison with the calculation of the excitation rate, and one should solve this problem first.

Analytical and numerical approaches for solving the problem of the rate of spontaneous emission of a single molecule located near the hole in a perfectly conducting and infinitely thin screen are presented in [23, 24].

However, in real experiments an approximation of a real metal film with a perfectly conducting and infinitely thin screen is not good enough, and one should take into account a finite conductivity of the metal and the finite thickness of the metal film with a hole. Taking these factors into account leads to the excitation of surface plasmon at the interface of the metal film and the dielectric and to Joule losses inside a metal.

In this work, we present a detailed study of the influence of the hole in the metal film with a finite thickness and conductivity on the rate of spontaneous emission of molecule located near it. As a model of a molecule, a classical oscillating dipole is used. Within this approximation, the decay rate can be found within well-known approaches [25 – 29]. Unfortunately, in the case of a real metal, there are no analytical solutions, and we have considered this problem numerically. Where it has been possible, however, we have used asymptotic expressions.

The rest of the paper is organized as follows. Section II describes the method of numerical simulation. Section III presents the spontaneous emission rate of a molecule with an arbitrary orientation of the dipole moment placed near the hole in the metal film in vacuum. In Section IV, we will investigate the spontaneous decay rate of the molecule near the hole in the metal film on a dielectric substrate.

## II. DESCRIPTION OF THE NUMERICAL SIMULATION

The problem of spontaneous emission of a molecule near the hole in the metal film with a finite thickness and conductivity has no analytical solution in general. Therefore, to solve this problem we make use of the numerical approach which is based on a software package COMSOL Multiphysics®.

The total spontaneous decay rate of a molecule located near the nanoaperture in the real metal film ($\gamma_{\text{total}}$), consists of a radiative ($\gamma_{\text{rad}}$) and non-radiative ($\gamma_{\text{nonrad}}$) rates of spontaneous emission (see Fig. 1).

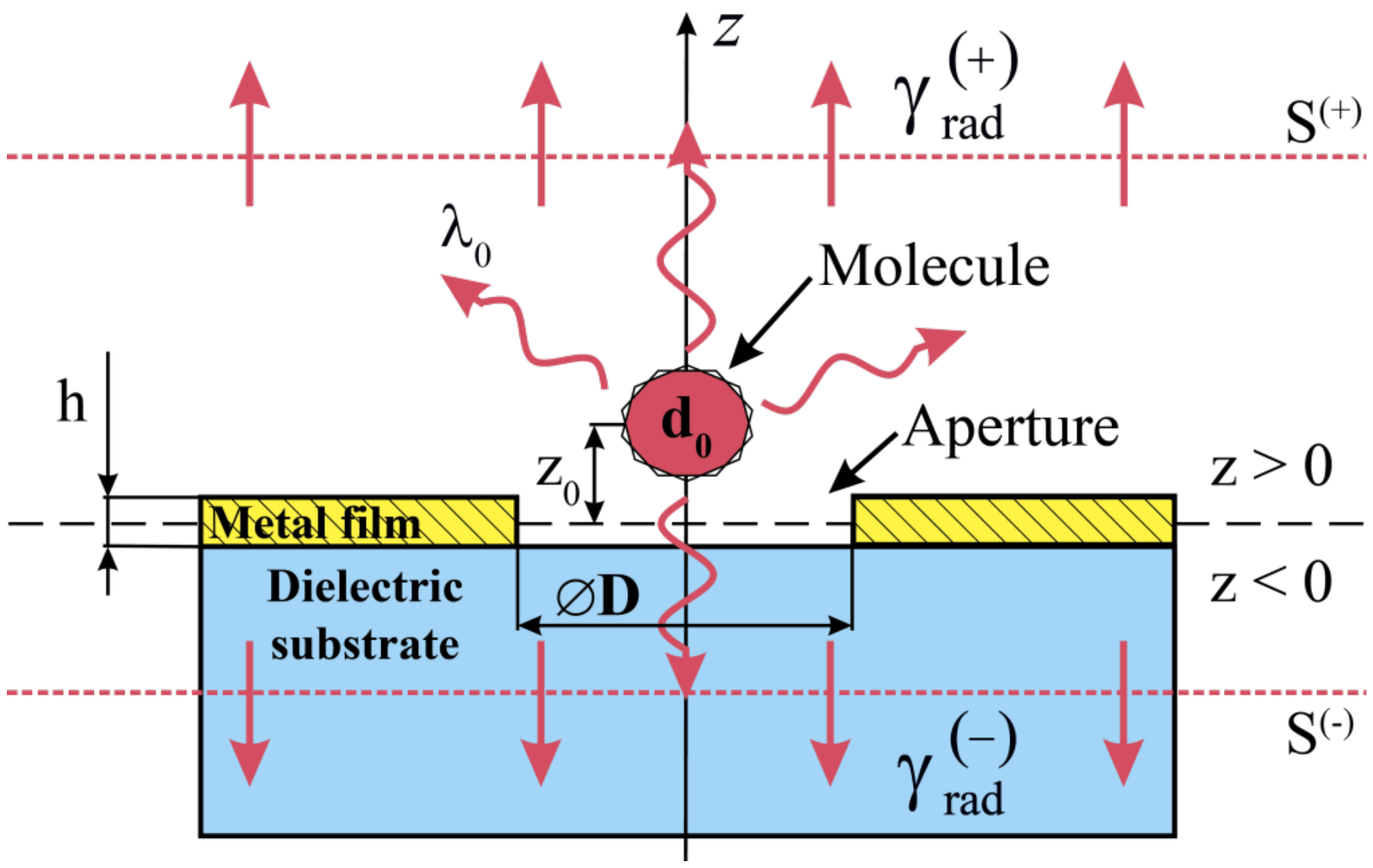

FIG. 1. Schematic representation of the geometry of the numerical simulation of the molecule radiation near the circular aperture of diameter *D* in the metal film of thickness *h*.

In its turn, the radiative decay rate ($\gamma_{\text{rad}}$) consists of radiative rates in the upper ($\gamma_{\text{rad}}^{(+)}$) and lower ($\gamma_{\text{rad}}^{(-)}$) half-spaces. The radiative decay rate corresponds to the emission of a photon by a molecule, which goes to infinity in the upper or lower half-spaces, respectively. The non-radiative decay rate of the molecule corresponds to the emission of a photon which is absorbed in the metal film. This can be expressed by the following formula:

$$\frac{\gamma_{\text{total}}}{\gamma_0} = \frac{\gamma_{\text{rad}}}{\gamma_0} + \frac{\gamma_{\text{nonrad}}}{\gamma_0} = \left(\frac{\gamma_{\text{rad}}^{(+)}}{\gamma_0} + \frac{\gamma_{\text{rad}}^{(-)}}{\gamma_0}\right) + \frac{\gamma_{\text{nonrad}}}{\gamma_0}, \tag{1}$$

where $\gamma_0$ is the rate of spontaneous emission of the molecule in vacuum.

Let us take the case of a weak interaction of the molecule with the electromagnetic field scattered on the aperture. Here, the total spontaneous decay rate of the molecule ($\gamma_{\text{total}}$) in vacuum near the hole can be calculated with the help of the solution of the classical problem of the diffraction of electromagnetic field of the oscillating point dipole source (with an optical frequency $\omega_0$ and a dipole momentum $\mathbf{d}_0$) on the aperture in a real metal film. If the oscillating point dipole source is defined by current density [25]:

$$\mathbf{j} = -i\omega_0 \mathbf{d}_0 \delta(\mathbf{r} - \mathbf{r}_0) \exp(-i\omega_0 t), \tag{2}$$

where $\delta(\mathbf{r}-\mathbf{r}_0)$ is Dirac delta function, $\mathbf{r}_0$ is coordinate of the position of the dipole, then the total rate of spontaneous emission of the molecule can be represented by the expression [25 – 29]:

$$\frac{\gamma_{\text{total}}}{\gamma_0}=\frac{3}{2}\text{Im}\left\{\frac{\mathbf{d}_0\cdot\mathbf{E}(\mathbf{r}_0,\mathbf{r}_0,\omega_0)}{k_0^3|\mathbf{d}_0|^2}\right\}=1+\frac{3}{2}\text{Im}\left\{\frac{\mathbf{d}_0\cdot\mathbf{E}^{(1)}(\mathbf{r}_0,\mathbf{r}_0,\omega_0)}{k_0^3|\mathbf{d}_0|^2}\right\}, \tag{3}$$

where $\mathbf{E}(\mathbf{r},\mathbf{r}_0,\omega_0)$ is the electric field, which is found from solving a full system of the Maxwell's equations with the oscillating point dipole source (2) located near the aperture in real metal film; $\mathbf{E}^{(1)}(\mathbf{r}_0,\mathbf{r}_0,\omega_0)$ is the scattered part of the field; $\gamma_0=\left(4k_0^3/3\hbar\right)|\mathbf{d}_0|^2$ stands for the rate of spontaneous emission in vacuum; $k_0=\omega_0/c_0=2\pi/\lambda_0$, $c_0$ is the speed of light in vacuum, $\lambda_0$ is the wavelength of the electric field in vacuum; Im describes the imaginary part.

Expression (3) describes the total spontaneous decay rate of the molecule, i.e. it takes into account the processes of absorption of the emitted photon by metal film, and the processes of real radiation where the photon goes to infinity.

The radiative part of the rate of spontaneous emission ($\gamma_{\text{rad}}$) can be expressed in terms of the flow of energy to infinity [25]:

$$\frac{\gamma_{\text{rad}}}{\gamma_0}=\frac{\gamma_{\text{rad}}^{(+)}}{\gamma_0}+\frac{\gamma_{\text{rad}}^{(-)}}{\gamma_0}=\frac{3}{8\pi k_0^4|\mathbf{d}_0|^2}\text{Re}\left\{\int_{S^{(+)}\cup S^{(-)}}dS\left(\mathbf{E}(\mathbf{r},\mathbf{r}_0,\omega_0)\times\mathbf{H}^*(\mathbf{r},\mathbf{r}_0,\omega_0)\right)_z\right\}, \tag{4}$$

where $S^{(+)}$ and $S^{(-)}$ are infinitely distant surfaces (see Fig. 1), $\mathbf{E}(\mathbf{r},\mathbf{r}_0,\omega_0)$ and $\mathbf{H}(\mathbf{r},\mathbf{r}_0,\omega_0)$ are the electric and magnetic fields at the observation point **r**, the asterisk denotes the complex conjugation, subscript *z* describes *z* component of the vector, Re describes the real part.

Non-radiative decay channel ($\gamma_{\text{nonrad}}$) corresponds to the absorption of photons in a metal film which results in metal heating. It is calculated by the volume integration:

$$\frac{\gamma_{\text{nonrad}}}{\gamma_0}=\frac{3\,\text{Im}\{\varepsilon_{\text{film}}(\omega_0)\}}{8\pi}\frac{\int_V|\mathbf{E}(\mathbf{r},\mathbf{r}_0,\omega_0)|^2\,dV}{k_0^3|\mathbf{d}_0|^2}, \tag{5}$$

where $\varepsilon_{\text{film}}(\omega_0)=\varepsilon'_{\text{film}}(\omega_0)+i\varepsilon''_{\text{film}}(\omega_0)$ is permittivity of the film material and *V* is the film volume.

We have investigated the metal film both (i) suspended freely in vacuum (see Section III) and (ii) lying on an infinite dielectric substrate ($\varepsilon_{\text{sub}}$ = 2.25) (see Section IV). In our simulation, the diameter of the hole *D* ranges from 50 to 1000 nm. The *z*-axis coincides with the axis of symmetry of the aperture. Coordinate *z* = 0 nm is the middle of the metal film. In case (ii), the permittivity of the upper half-space and the hole is assumed to be equal to one (vacuum). In our approach, the molecule is described by a point oscillating electric dipole, which is defined by current density (2). It is located on the symmetry axis of the aperture and has the coordinate $z_0$. The electric dipole with the dipole moment $\mathbf{d}_0$ radiates in vacuum on the wavelength $\lambda_0$ = 500 nm. In the numerical simulation, we use the optical properties of the gold metal film from [30] (for $\lambda_0$ = 500 nm the permittivity of gold is –2.13 + 2.42$i$ [30]). In the case of a vertical dipole, the dipole moment $\mathbf{d}_0$ has the same direction as the $z$-axis, in the case of the horizontal dipole $\mathbf{d}_0$ is perpendicular to the $z$-axis.

In our numerical simulation, we use the finite element method (its commercial product realization in COMSOL Multiphysics®) for solving Maxwell's equations with a point oscillating electric dipole source defined by current density (2).

## III. MOLECULE LOCATED NEAR A SINGLE APERTURE IN THE METAL FILM SUSPENDED IN VACUUM

Let us consider the rate of spontaneous emission of a molecule located near a single aperture in the metal film suspended in vacuum. There are two variants for orientation of the dipole moment of the molecule transition: vertical (i.e. $\mathbf{d}_0 \parallel z$) and horizontal (i.e. $\mathbf{d}_0 \perp z$). Fig. 2 illustrates the total rate of spontaneous emission of the molecule as a function of the hole diameter $D$ and position of the molecule on the symmetry axis $z_0$. The radiation wavelength $\lambda_0$ is 500 nm.

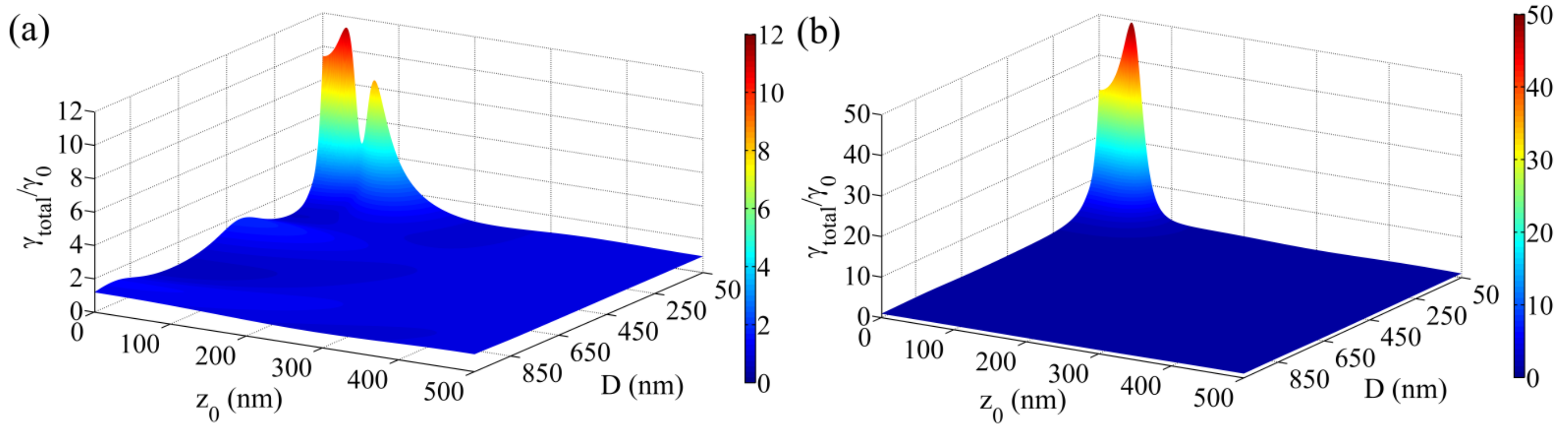

FIG. 2. The total rate of spontaneous emission of the molecule ($\gamma_{total}/\gamma_0$) located on the axis of the aperture in the gold film as a function of its position $z_0$ and the hole diameter $D$. The radiation wavelength $\lambda_0$ is 500 nm. The thickness of the metal film $h$ is 100 nm. (a) The vertical orientation of the dipole moment, (b) the horizontal orientation of the dipole moment.

Figure 2(a) shows that for vertical orientation of the dipole moment and the range of aperture diameter 50 – 100 nm the total rate of spontaneous emission of the molecule has two clearly pronounced maximums. The large maximum is inside the aperture, the smaller is above the film surface. The minimum is near the boundary of the metal film. By increasing the hole diameter or the molecule position $z_0$, the total rate of spontaneous emission tends to one as in the case of emission of the molecule in vacuum.

Figure 2(b) illustrates that for the horizontal dipole moment, in contrast to the vertical one, for small values of the diameter and near the aperture the total rate of the radiation has only one maximum. This peak is inside the aperture. At this maximum, the total rate is four times greater than in the case of vertical orientation of the dipole moment. Also, Fig. 2(b) shows that the total rate of spontaneous emission tends to one as in the case of the spontaneous emission of the molecule in vacuum when the hole diameter or the position coordinate of the molecule $z_0$ is increased. Note that in the domain of the small diameter and the small values of the position coordinate, the main contribution to total rate of radiation is the non-radiative channel.

Figure 3 illustrates the total rate of the spontaneous emission of the molecule as a function of the molecule position $z_0$ and the emission wavelength $\lambda_0$ (the diameter of the aperture is 100 nm).

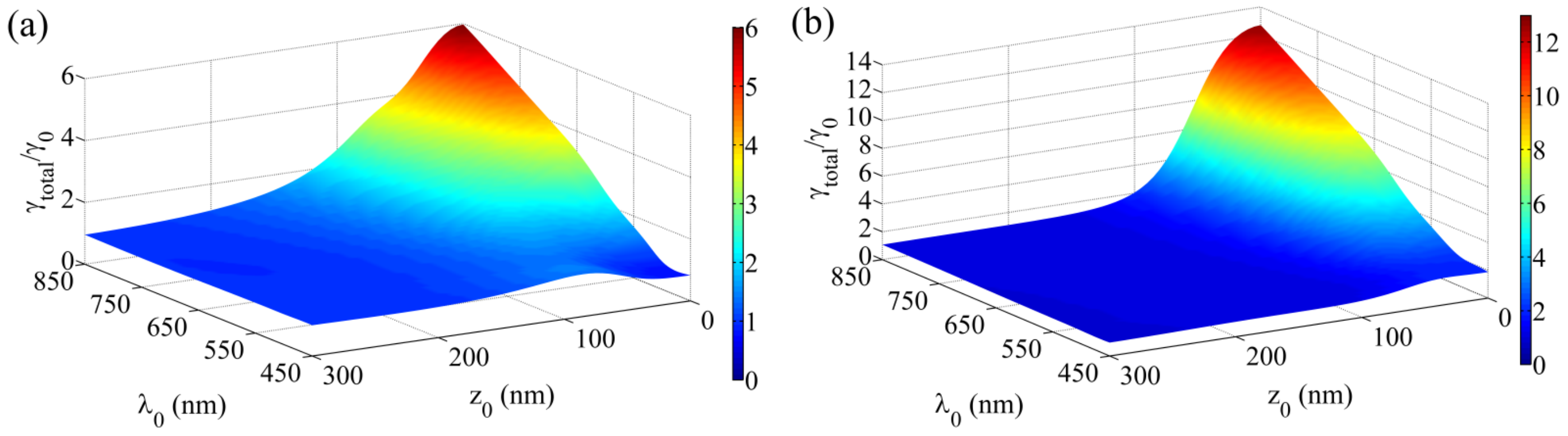

FIG. 3. The total rate of spontaneous emission of the molecule ($\gamma_{total}/\gamma_0$) is located on the axis of the aperture as a function of its position $z_0$ and emission wavelength $\lambda_0$. The diameter of the aperture is 100 nm. The thickness of the metal film *h* is 100 nm. (a) The vertical orientation of the dipole moment, (b) the horizontal orientation of the dipole moment.

Figures 3(a) and 3(b) show that for vertical and horizontal orientation of the dipole moment in the considered range of the radiation wavelengths and the positions of the molecule dependence generally has a monotonous character. This indicates a non-resonant character of the interaction of the molecule with the hole. The total rate increases with the increasing wavelength radiation and the decreasing distance of the molecule to the hole.

Figure 4 shows the radiative rate of spontaneous emission of the molecule in the upper half-space as a function of the position of the molecule ($z_0$ varies from –500 to 500 nm) and the hole diameter (*D* ranges from 50 to 1000 nm). Usually, it is this rate that determines the observable flux of photons.

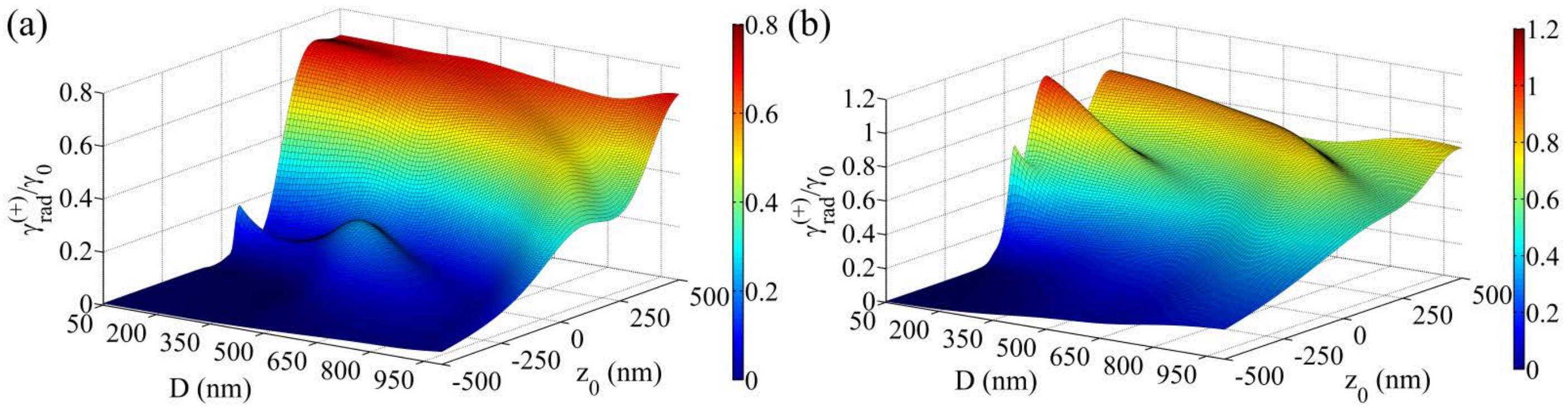

FIG. 4. The radiative rate of spontaneous emission in the upper half-space of the molecule ($\gamma_{rad}^{(+)}/\gamma_0$) is located on the axis of the aperture in the gold film as a function of its position $z_0$ and the diameter of the hole *D*. The radiation wavelength $\lambda_0$ is 500 nm. The thickness of the metal film *h* is 100 nm. (a) The vertical orientation of the dipole moment, (b) the horizontal orientation of the dipole moment.

Figure 4 shows that in the cases of vertical and horizontal orientations of the dipole moment radiative rate in the upper half-space has the minimum for the negative values $z_0$ and the small values of the hole diameter. It rises with the increase of these parameters. When the position of the molecule is in the lower half-space, the radiative rate in the upper half-space decreases very quickly. Comparing Fig. 2 and Fig. 4, one can see that the radiative rate is a small part of the total spontaneous emission rate. Therefore, the non-radiative spontaneous decay rate gives the main contribution to the total decay rate of the molecule. This fact is more pronounced if the hole diameter is small and the molecule is close to the aperture.

For a clearer understanding of the relations between the total, radiative, and non-radiative decay rates of the molecule spontaneous emission in Fig. 5, we show their dependencies on the molecule position $z_0$ for the aperture diameter 100 nm.

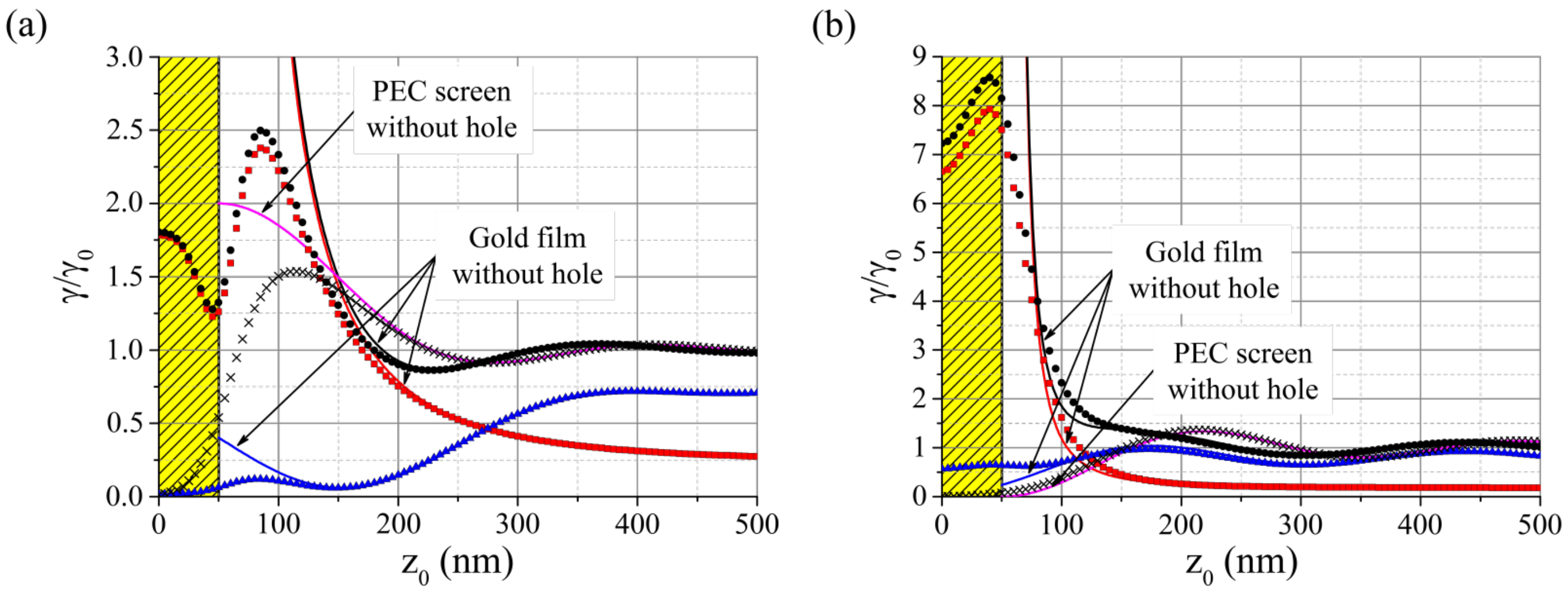

FIG. 5. Total $\gamma_{total}/\gamma_0$ (black color), radiative $\gamma_{rad}/\gamma_0$ (blue color), non-radiative $\gamma_{nonrad}/\gamma_0$ (red color) rates of spontaneous emission of the molecule located on the axis of symmetry of the aperture in the gold film as a function of molecule position $z_0$. The black points, the blue triangles, the red squares are the case of the gold film with an aperture. The solid lines are the case of the gold film without a hole [27, 31]. The solid magenta line is the total rate of spontaneous emission for the perfect electric conducted (*PEC*) screen without a hole [24, 28, 29]. The black crosses are the total spontaneous rate of the molecule located near the PEC screen with the aperture (the diameter of the hole is 100 nm, the thickness of the film is 100 nm). The yellow stripe is the position of the metal or the PEC film. The wavelength $\lambda_0$ is 500 nm. The thickness of the metal film *h* is 100 nm. The diameter of the aperture *D* is 100 nm. $z$ = 0 nm corresponds to a plane passing through the middle of the metal film. $\gamma_0$ is the total rate of spontaneous emission of the molecule in vacuum. (a) The vertical orientation of the dipole moment, (b) the horizontal orientation of the dipole moment.

From Fig. 5(a) for the vertical orientation of the dipole moment, it can be seen that in the presence of finite conductivity non-radiative contribution becomes very significant, or even predominant. The radiation contribution begins to prevail over the non-radiative one only at very large distances, when the coordinate $z_0$ is greater than 3*D*. It can be seen that, when $z_0$ is greater than 2*D*, the molecule ceases to feel the presence of the hole, and for the calculation of total, radiative, non-radiative rates the approach «molecule located near the metal plane» [27, 31] starts to work well. Note that for a large enough distance between the dipole and the metal surface, the total rate of spontaneous emission of the molecule tends to the asymptote «molecule near a perfectly conducting infinitely thin screen without a hole» [24, 28, 29].

From Fig. 5(b), it can be seen that even at a distance only slightly greater than *D*, the radiative emission channel begins to prevail over the non-radiative channel. When the molecule is close to the surface of the metal ($z_0 < D$), the significant contribution to the total spontaneous decay rate is due to the non-radiative channel. If the molecule remoteness is more than *D*, the influence of the hole on the molecular emission becomes weak, and the approach «the molecule near the metal film without a hole» stats to work well [27, 31]. At a distance of more than 4*D*, the total spontaneous decay rate of the molecule can be described right by the asymptotic «molecule near a perfectly conducting screen without a hole» [24, 28, 29].

It can be seen that the non-radiative channel, which is defined by the metal heat absorption and excitation of the surface plasmon, gives a significant contribution to the total spontaneous emission rate of the molecule near the nanohole. This contribution is especially essential for the horizontal orientation of the dipole moment.

In the case of the metal film suspended freely in vacuum, only the «ordinary» surface plasmon wave that propagates at the boundary of «metal-dielectric» can be excited. This surface plasmon wave is converted into heat and could not be re-emitted into the surrounding space. In this sense, the separation of the contribution of the plasmon and non-radiative losses becomes impossible.

## IV. A MOLECULE LOCATED NEAR THE SINGLE APERTURE IN THE METAL FILM PLACED ON A DIELECTRIC SUBSTRATE

Let us consider the rate of spontaneous emission of the molecule located near a single hole in a metal film placed on the semi-infinite dielectric substrate ($\varepsilon_{\text{sub}}$ = 2.25). In this section, the thickness of the metal layer $h$ is 80 nm. We have chosen such thickness for comparison with the study of radiative decay rates enhancement due to «leaky» surface plasmon waves in a multilayer structure without a hole [31]. Fig. 6 shows the dependence of the total, radiative, and non-radiative of the decay rate as a function of the molecule position on the axis of the hole and the orientation of its dipole moment.

The comparison of Fig. 5(a) and Fig. 6(a) shows that in the cases of a gold film with the thickness 100 nm in vacuum and the thickness 80 nm on the substrate, the pattern of the spontaneous decay rate remains almost unchanged. This is primarily due to the great absorption of gold.

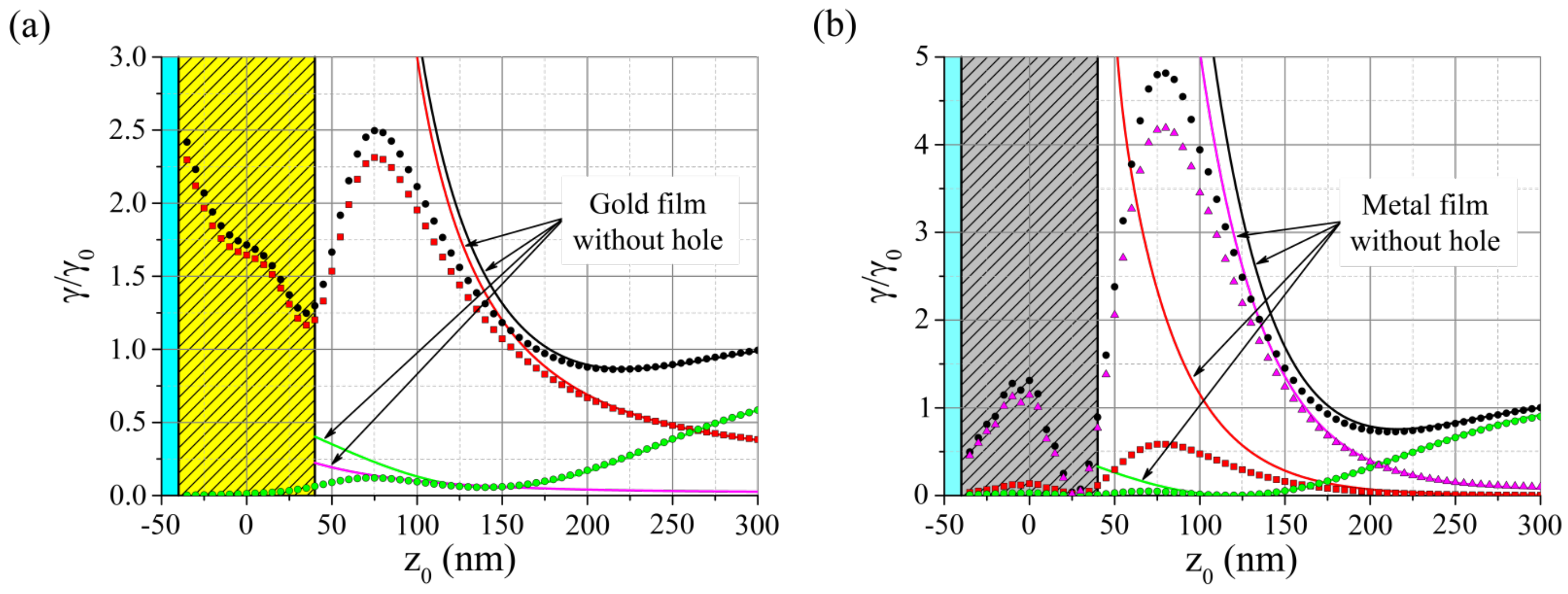

FIG. 6. Total $\gamma_{\text{total}}/\gamma_0$ (black color), radiative into the upper $\gamma_{\text{rad}}^{(+)}/\gamma_0$ (green color) and into the lower $\gamma_{\text{rad}}^{(-)}/\gamma_0$ (magenta color) half-spaces, non-radiative $\gamma_{\text{nonrad}}/\gamma_0$ (red color) rates of spontaneous emission of the molecule located on the axis of symmetry of the aperture in the metal film as a function of the position of the molecule $z_0$. The points are the case of the metal film with an aperture. The solid lines are the case of the metal film without a hole [27, 31]. The yellow or gray stripe is the position of the metal film. The light blue stripe is the position of the dielectric substrate ($\varepsilon_{\text{sub}}$ = 2.25). The wavelength $\lambda_0$ is 500 nm. The thickness of the metal film *h* is 80 nm. The diameter of the aperture *D* is 100 nm. *z* = 0 nm corresponds to a plane passing through the middle of the metal film. $\gamma_0$ is the total rate of spontaneous emission of the molecule in vacuum. There is the vertical orientation of the dipole moment. (a) The gold film (for $\lambda_0$ = 500 nm $\varepsilon_{\text{film}}$ = – 2.13 + 2.42*i*) [30], (b) the film of a hypothetical metal material (for $\lambda_0$ = 500 nm $\varepsilon_{\text{film}}$ = –1.72 + *i*0.001).

However, if there is a dielectric substrate, the excitation of «leaky» surface plasmon wave becomes possible [31, 32]. This leads to the fact that at certain parameters of the system it is possible to obtain a substantial increase in the spontaneous radiative decay rate into the lower half-space. To excite the «leaky» surface plasmon wave in the considered system, one need to take a specially selected metal film (for $\lambda_0$ = 500 nm the permittivity of the metal film $\varepsilon_{\text{film}}$ is –1.72 + *i*0.001). The results of calculations for the latter case are shown in Fig. 6(b). It can be seen that in this case, the radiative rate in the lower half-space starts to make a prevailing contribution to the rate of spontaneous decay.

Fig. 7 shows how this effect depends on the magnitude of the imaginary part of the permittivity of the metal film (Fig. 7(a) $\varepsilon_{\text{film}}$ = –1.72 + *i*0.01, Fig. 7(b) $\varepsilon_{\text{film}}$ = –1.72 + *i*0.1).

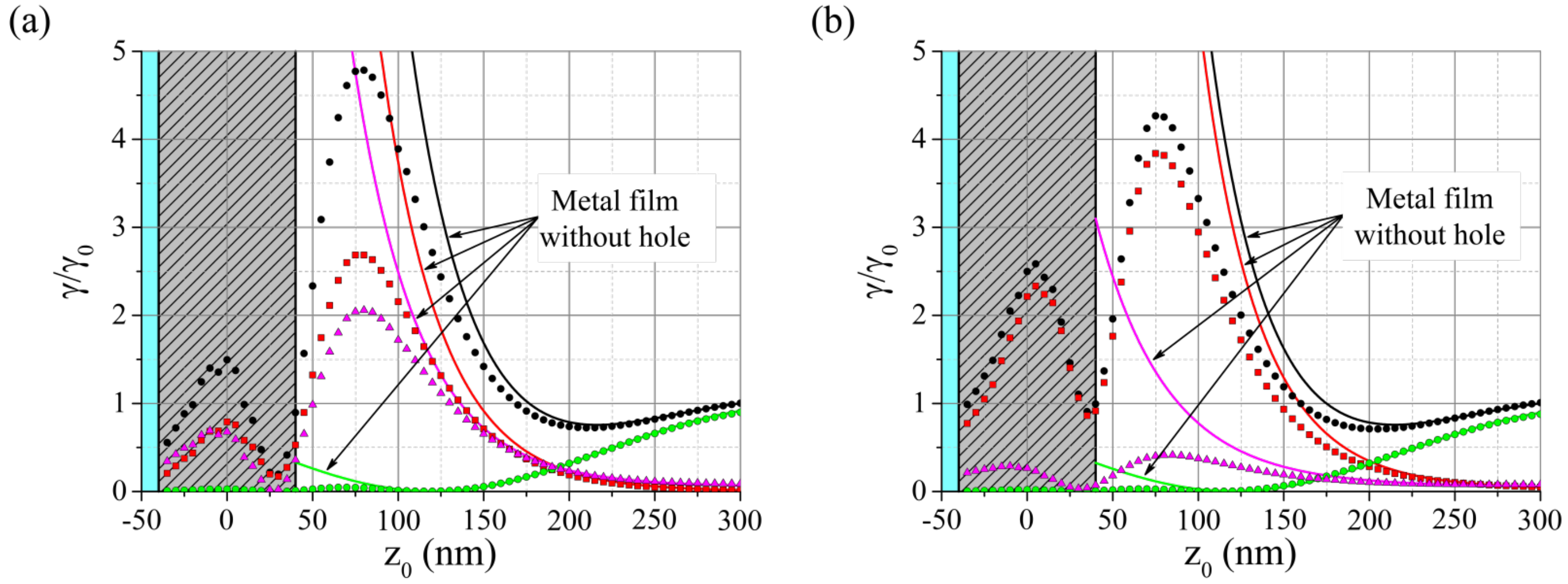

FIG. 7. Total $\gamma_{\text{total}}/\gamma_0$ (black color), radiative into the upper $\gamma^{(+)}_{\text{rad}}/\gamma_0$ (green color) and into the lower $\gamma^{(-)}_{\text{rad}}/\gamma_0$ (magenta color) half-spaces, non-radiative $\gamma_{\text{nonrad}}/\gamma_0$ (red color) rates of spontaneous emission of the molecule located on the axis of symmetry of the aperture in a metal film as a function of the molecule position $z_0$. The points are the case of the metal film with an aperture. The solid lines are the case of the metal film without a hole [27, 31]. The gray stripe is the position of the metal film. The light blue stripe is a position of the dielectric substrate ($\varepsilon_{\text{sub}}$ = 2.25). The wavelength $\lambda_0$ is 500 nm. The thickness of the metal film *h* is 80 nm. The diameter of the aperture *D* is 100 nm. *z* = 0 nm corresponds to a plane passing through the middle of the metal film. $\gamma_0$ is the total rate of the spontaneous emission of the molecule in vacuum. There is the vertical orientation of the dipole moment. There is a film of a special metal material (for $\lambda_0$ = 500 nm (a) $\varepsilon_{\text{film}}$ = –1.72 + *i*0.01, (b) $\varepsilon_{\text{film}}$ = –1.72 + *i*0.1).

The comparison of Fig. 6(b) and Fig. 7 shows that the intensity of the effect of increasing radiative rate in the lower half-space depends essentially on the imaginary part of the metal film. If the value of the imaginary part is 0.1, the effect becomes insignificant, and the non-radiative channel becomes dominant. This effect is observed only at small losses ($\varepsilon''_{\text{film}}$ < 0.1). Such low losses are difficult to implement in an experiment directly. Nevertheless, to check these result, one can use the gallium-doped zinc oxide as an example (it has $\varepsilon_{\text{film}}$ ≈ –1.05 + *i*0.6 [33]) and a permittivity of the substrate of $\varepsilon_{\text{sub}}$ ≈ 10 (which is the case of semiconductors in the infrared range of the wavelength). For these materials in the case without a hole, one can obtain a radiative decay enhancement in the lower half space equal to 8 [31], which exceeds substantially the radiative rate achievable in other configurations considered in this paper. Even better results can be achieved with dysprosium-doped cadmium oxide at wavelength around 1950 nm ($\varepsilon''_{\text{film}}$ ≈ 0.2 [34]) or with crystalline silicon carbide in the mid-infrared ($\varepsilon''_{\text{film}}$ ≈ 0.15 [35]). For these materials in the case without a hole, one can expect the radiative decay rate to be enhanced 30-fold with the same substrate ($\varepsilon_{\text{sub}}$ ≈ 10) [31]. Moreover, the concept of loss compensation is currently widely discussed [36] and can be used to achieve the conditions of realization of the found effect.

The presence of the substrate can change the relation between the radiative and non-radiative channels of the spontaneous emission radically. Especially clearly, this effect is manifested in the excitation of «leaky» surface plasmon wave in this system.

## V. CONCLUSION

In this paper we have investigated numerically the spontaneous emission of a single molecule located near the hole in the finite thickness gold film. We have made the comparative analysis of two cases: an aperture in a perfectly conducting finite thickness film and a hole in a metal film. The essential influence of the molecule position on the rate of spontaneous emission is shown. We have pointed out the cases where asymptotes can be used to describe this process. Visual relation to the total, radiative and non-radiative rates of spontaneous emission of the molecule is presented.

It is shown that for the considered parameters, an approximation model of the total spontaneous rate of the molecule «molecule located near a perfectly conducting screen» starts to work well only at large enough distances ($z_0 > 4D$) of the molecule from the aperture.

It is also demonstrated that the presence of the substrate and the occurrence of a «leaky» surface plasmon wave change considerably the relation between the radiative and non-radiative decay channels. They increase significantly the radiative rate in the lower half-space. This effect is observed only at relatively small losses which are difficult to implement in an experiment directly. Nevertheless, to check these result one can use the gallium-doped zinc oxide as an example (it has $\varepsilon_{\mathrm{film}} \approx -1.05 + i0.6$ [33]) and a permittivity of the substrate of $\varepsilon_{\mathrm{sub}} \approx 10$ (which is the case of semiconductors in the infrared range of the wavelength). For these materials in the case without a hole, one can obtain a radiative decay enhancement in the lower half-space equal to 8 [31], which substantially exceeds the radiative rate achievable in other configurations considered in this paper. Even better results can be achieved with dysprosium-doped cadmium oxide at wavelength around 1950 nm ($\varepsilon''_{\mathrm{film}} \approx 0.2$ [34]) or with crystalline silicon carbide in the mid-infrared ($\varepsilon''_{\mathrm{film}} \approx 0.15$ [35]). For these materials in the case without hole, one can expect the radiative decay rate to be enhanced 30-fold with the same substrate ($\varepsilon_{\mathrm{sub}} \approx 10$) [31]. Moreover, the concept of loss compensation is currently widely discussed [36] and can be used to achieve the conditions of realization of the found effect.

In this paper, we consider only the spontaneous emission of the molecule near the hole in the metal film. The complete process of a fluorescence molecule near a nanohole will be considered in a separate publication [37].

**ACKNOWLEDGEMENTS**

I. V. Treshin thanks Alexey B. Novikov and Konstantin V. Ivanov for advance help in high performance calculation.

The research has been supported by the Advanced Research Foundation (Contract No. 7/004/2013-2018). The authors are also grateful to the Russian Foundation for Basic Research (Grants No. 14-02-00290 and No. 15-52-52006).